\documentclass[reprint,amsmath, amssymb, aps, floats]{revtex4-1}

\usepackage{graphicx}

\begin{document}


\title{The effect of a scanning flat fold mirror on a CMB B-mode experiment}

\author{William F. Grainger, Chris E. North, Peter. A. R. Ade}
\affiliation{Astronomy Instrumentation Group, Cardiff University.}

\date{\today}


\begin{abstract}
We investigate the possibility of using a flat-fold beam steering mirror
for a CMB B-mode experiment. An aluminium flat-fold mirror is found to
add $\sim$0.075\% polarization, which varies in a scan synchronous
way. Time-domain simulations of a realistic scanning pattern are
performed, and the effect on the power-spectrum illustrated and
a possible method of correction applied.  
\end{abstract}

\pacs{1}

\maketitle

\section{Introduction}

The prevailing $\Lambda$CDM cosmological model has been very
successful in explaining many observations of the universe. However,
inflation - exponential expansion within the first $\sim 10^{-35}$~s - 
has many uncertain details; indeed it lacks strong
confirmation. Inflationary models predict a background of gravity
waves \cite{Starobinsky1, Starobinsky2, Rubakov, Grishchuk,
Abbott_Wise}
which produce a curl-like, or ``B-mode'' pattern of polarization on
the cosmic microwave background (CMB) on scales greater than
$1^{\circ}$, which cannot be produced by density perturbations
\cite{WhyB, ZandS}. The amplitude of this signal is characterized by
the tensor-to-scalar ratio, $r$, which is constrained by current
measurements of the temperature power spectrum and
temperature-polarization cross-correlation to a value $r<0.20$
\cite{wmap_7year}. To put tight constraints on inflationary models
\cite{taskforce} we need to detect the primordial B-mode signal, or
constrain it for $r<0.01$. This requires measurement of polarization
fluctuations at the level of 10~nK.


To measure this signal level is challenging, and good control of
systematics is required \cite{Hu_Matias_params}. A number of
ground-based (QUIET \cite{QUIET}, PolarBear \cite{PolarBear}),
balloon-based (EBEX \cite{EBEX_Grainger_SPIE}, Spider
\cite{Spider_SPIE}) and satellite-based (Planck
\cite{PlanckLFI,PlanckHFI}) experiments have been, or are close to
being, fielded with the raw sensitivity to detect the B-mode signal if
it is present at a reasonable $r$. These experiments are all designed
to have low, and stable, instrumental polarization (IP), and often
include a modulation strategy to assist with reducing systematics and
mitigating against $1/f$ noise. The majority of these experiments hold
their optics stationary with respect to the receiver, and scan the
cryostat and optics together; a notable exception being the Tenerife experiment
\cite{Tenerife}. Scanning the cryostat places requirements on the
stability with respect to accelerations and tilts. 
Although cryogenic receivers have been demonstrated that are stable at
different tilt angles \cite{grainger_clover_spie}, the sensitivity to
accelerations has not been demonstrated. A system can be imagined,
however, which combines a stationary receiver with a fold
mirror. Besides the simplification of the cryogenic requirements, the
large reduction in the mass of moving components would make the design
of the steerable mount much simpler.

We investigate such a system in this paper. In
Section~\ref{sec:instrument} we describe the system, and review the
production of polarization from reflections. In Section~\ref{sec:sims}
we describe a fiducal scan strategy and simulation pipeline. Results
from this pipeline are then presented and discussed, along with other
potential problems in Section~\ref{sec:problems}.

\section{Method}
\label{sec:instrument}

For this investigation, we consider a horn fed receiver and compact
range antenna (CRA), placed in a groundscreen, nominally observing the
horizon.  The output beam of the CRA is then steered to the sky with a
single, flat, fold mirror. Although the flat mirror would be large;
$\sim 4 \times 4$~m, such a system can be considered. The flat mirror is then
tipped and tilted such that the beam of the telescope moves across the
sky to allow observations.

Off-normal reflections, as caused by this folding, induce a
polarization. The effect is well known and is the principle behind,
for example, the ``pile-of-plates polarizer''. The Fresnel
equations give the reflection coefficients for incident
angle $\theta_i$ and transmission angle $\theta_t$ for an incident
wave with E-field parallel to the boundary, $R_{\parallel}$ and
$R_{\perp}$:
\begin{equation}
R_{\parallel} =
\left(\frac{\tan^2(\theta_i-\theta_t)}{\tan^2(\theta_i+\theta_t)}
\right)
\end{equation}
and
\begin{equation}
R_{\perp} =
\left(\frac{\sin^2(\theta_i-\theta_t)}{\sin^2(\theta_i+\theta_t)}
\right)
\end{equation}
For a metal plate, we use the relation for the complex index of refraction,
\begin{equation}
(n+i\kappa)^2=\epsilon' + i\epsilon''
\end{equation}
For a general dielectric, the conductivity $\sigma$ is given by
$\sigma = \omega \epsilon'' \epsilon_{0}$ where $\omega = 2 \pi
f$ and $f$ is the radiation frequency. For a good conductor, we can
assume $\epsilon' \ll \epsilon''$, and so
\begin{equation}
n^2=\frac{\sigma}{2\omega \epsilon_{0}}
\end{equation}
Plugging this into Snell's law allows calculation of $R_{\parallel}$
and $R_{\perp}$. For a pure aluminium plate, $1 / \sigma = \rho = 28.2
\times 10^{-9}$~$\Omega {\rm m}$, we find $R_{\parallel} = 0.9970$ and
$R_{\perp} = 0.9985$. Thus an unpolarized input reflected with
$\theta_i = 45$ would have an overall polarization of $0.075\%$
This ignores the phase-delay from the reflections, which 
elliptically polarizes the reflected beam, producing both
circular polarization and rotating the plane of polarization. Using a
Drude-Lorentz model to calculate $n$ and $\kappa$ for aluminium at
90GHz\cite{DLmodel}, we calculate that the plane of polarization is rotated by an
additional $\sim 10^{-9}$ degrees when the phase delay is
included. Ignoring this is then valid as this is a much smaller than
the effect from differential reflection ($\sim 0.2^{\circ}$).
The reflection is a total power to polarization conversion ($T \to
Q,U$); in the context of a CMB polarization measurement the total power
is 2.7~K, and so the induced polarized signal is then $\sim$200~$\mu {\rm
K}$, higher than even the EE polarization spectrum. However, the
signal is changing consistently with the scan, and so is varying on
different scales from the CMB signals of interest and so could, in
principle, be removed. In the next section, we describe the pipeline
with which we investigate the effect on CMB reconstruction and
$C_{\ell}$ estimation.

\begin{figure*} 
\begin{center}
\includegraphics[width=5in]{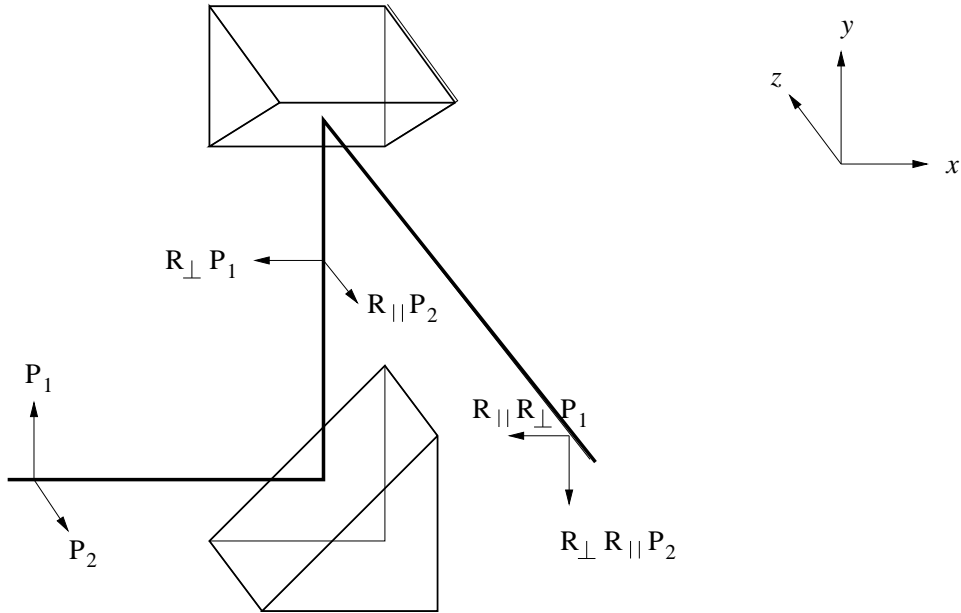}
\caption{A pair of flat-fold mirrors with no net polarization.}
\label{fig:nonetpolarization}
\end{center}
\end{figure*}

We note in passing that it is possible to construct a two mirror
system that provides a fold with no net polarization. Consider a beam
travelling along the $x$ axis. The first mirror, mounted at
45$^{\circ}$ to the $yz$ plane, then reflects the beam so it is
travelling along the $y$ axis, and the second, mounted at 45$^{\circ}$
to the $xy$ plane, reflects that beam along the $z$. As the reflection
axes have switched, both input $E_x$ and $E_y$ receive an attenuating
factor of $R_{\parallel} R_{\perp}$.  Any net polarization is then due
to differences between the conductivity of the mirrors. If the two
mirrors are fixed relative to each other, and allowed to rotate about
the $x$ axis, the beam can be steered. A second set of mirrors, in a
similar configuration, can be added to allow full beam steering around
the sky. This is illustrated in figure~\ref{fig:nonetpolarization}. 

\section{Scan strategy and pipeline}
\label{sec:sims}

We simulate observations with the following four step pipeline.
Firstly, time ordered data (TOD) is simulated with 100Hz sampling for
four 3-hour observation periods, corresponding to a day of
observations, and the sky location (RA, Dec), telescope pointing
direction (az, el), and observed $E_x^2$ and $E_y^2$ recorded.  The
input CMB maps were produced by simulating a power spectrum with CAMB
\cite{camb}, and producing a realization with the
Healpix\footnote{http://healpix.jpl.nasa.gov} \cite{healpix} {\em
  synfast} routine. This input all-sky map was also convolved with a
Gaussian beam with FWHM of 5.5 arcmin. 
Parameters used in the simulation are listed in
Table~\ref{tab:simparams1}. Note that we include the monopole in this
simulation, with a temperature of 2.7~K and so $E_x^2 \sim E_y^2 \sim= 1.3$~K.

It is generally desirable for ground-based and suborbital experiments
to scan the sky at constant elevation to minimize the effect of
variable airmass. We follow this and consider the proposed Clover scan
strategy, the optimization of which is fully described elsewhere
\cite{cen_brad_clover_strat}. The strategy is to perform azimuthal
scans at fixed elevation (nominally $45^{\circ}$) as the chosen sky
patch rises for around 1 hour, continually adjusting the azimuthal
center to achieve as circular a patch as possible. Once the patch has
risen sufficiently through the scan, the elevation is increased,
nominally to $\sim60^{\circ}$, and the patch continues to rise for
another 3 hours. The same strategy is used as the patch sets, giving
four distinct data streams for a single patch during a 8-12-hour
observation period.

The second step in the pipeline is to take the pointing from the TOD
data, determine the angle of the flat folding mirror, rotate the
observed $E_x^2$ and $E_y^2$ to align with the plane of the mirror,
apply the calculated $R_{\parallel}$ and $R_{\perp}$, and then rotate
back.  The TOD is then naively binned to produce T, Q and U maps,
following the map-making scheme outlined in
\cite{Brown_clover_modulation}, and then written out as a Healpix
format FITS files. We make maps at $N_{\rm side}=512$ so that the
multipole range $\ell \sim 100$ to $1000$ is retrievable,
and avoid holes in the maps by simulating a five pixel focal
plane. The pixel angular positions from boresight are listed in
Table~\ref{tab:simparams1}. 

Finally, the angular power spectrum (APS) coefficients $C_\ell$
are estimated using the {\em anafast} Healpix routine. We apply
a circular mask of radius $10^{\circ}$ with a simple cosine-squared
taper of width $2^{\circ}$ at the edge.

\begin{table}
\caption{Simulation parameters. Detector positions are given in (az, el) in the telescope frame.}
\label{tab:simparams1}
\begin{ruledtabular}
\begin{tabular}{cc}
Parameter & Value \\
\colrule
Telescope location & Atacama desert\\
Patch centre & 9h00m,-44$^{\circ}$00' \\
Patch radius & 10$^{\circ}$ \\
Azimuthal scan speed & 0.5 deg/s \\
Nominal elevations & 45, 60 deg \\
Observing frequency & 90 GHz\\
Data rate & 100 Hz \\
Resitivity of aluminium & $28.2\times 10^{-9}$~$\Omega {\rm m}$ \\
Input {\em r} & 0.1 \\
Detector 1 position & (0,0) arcmin\\
Detector 2 position & (14.8, 14.8) arcmin\\
Detector 3 position & (29.6, 29.6) arcmin\\
Detector 4 position & (36.9, 7.4) arcmin\\
Detector 5 position & (51.7, 22.1) arcmin\\ 
\end{tabular}
\end{ruledtabular}
\end{table}

\subsection{Determination of angles}

\begin{figure*}
\begin{center}
\includegraphics[width=6in]{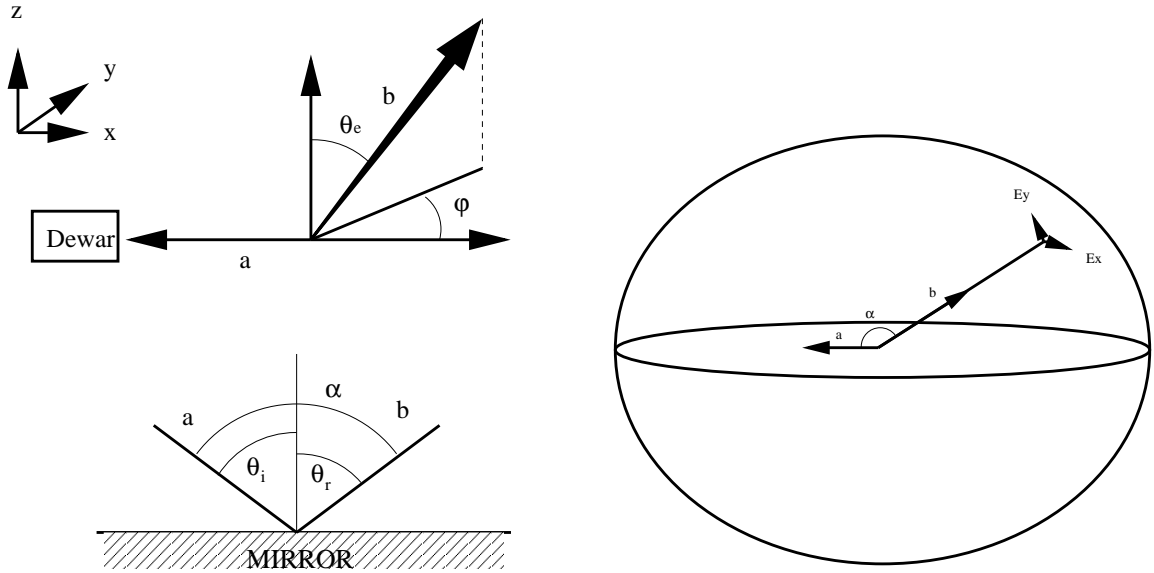}
\caption{Fiducial geometry considered for calculating the relevant
  angles. Top left: the telescope beam from the dewar (direction ${\bf a}$) to the sky (${\bf b}$), directed at the azimuthal polar coordinates ($\theta_e, \phi$). Bottom left: the angle $\alpha$ through which the beam is reflected by the mirror. Right: $E_x$ and $E_y$, the two polarization vectors on the sky.}
\label{fig:defineab}
\end{center}
\end{figure*}

In order to determine the effect for a mirror at an arbitrary orientation,
two angles are required; the angle of incidence, $\theta_i$ and the
angle, $\beta$, between the mirror and the horizon as seen by the
incident photons.  Consider the the flat fold mirror to be at the
origin, and the unit vector {\bf a} points to the dewar, and {\bf b}
points to the position on the sky being observed. This is is
illustrated in figure~\ref{fig:defineab}. $\alpha$ is the angle
between {\bf a} and {\bf b}. In polar coordinates, ${\bf a} = (1,
90^{\circ}, 180^{\circ})$ and ${\bf b} = (1, \theta_e, \phi)$ where
$90 - \theta_e$ and $\phi$ are the the elevation and azimuth of the
sky location. Converting into Cartesian coordinates, we have ${\bf a}
= (-1, 0, 0)$ and ${\bf b} = (\sin \theta_e \cos \phi, \sin \theta_e
\sin \phi, \cos \theta_e)$. To determine $\alpha$, we take the
dot-product of {\bf a} and {\bf b} and get
\begin{equation}
 {\bf a} . {\bf b} = \cos \alpha
                   = - \sin \theta_e \cos \phi
\end{equation}
As the angles of incidence and reflection, $\theta_i$ and $\theta_r$
are equal, we have 
\begin{equation}
 \theta_i = 1/2 \alpha
          = 1/2 \arccos (- \sin \theta_e \cos \phi))
\end{equation}

We define $E_x$ to be parallel to the horizon, and $E_y$ to be
``upwards'' towards the zenith. We require the angle $\beta$ by which
$E_x$ and $E_y$ need to be rotated (around the line of sight) so that
the rotated $E_x$ is parallel to the mirror. We define ${\bf c}$, a
unit vector that points in the same direction as $E_x$, so, in polar,
coordinates, $c = (1, 90^{\circ}, \phi+90^{\circ})$.  ${\bf a} \wedge
{\bf b}$ is parallel to the mirror. By taking the dot product of ${\bf
  a} \wedge {\bf b}$ with ${\bf c}$, we can determine the angle
$\beta$ as
\begin{equation}
  ({\bf a} \wedge {\bf b}) . {\bf c} = | ({\bf a} \wedge {\bf b}) | |
  {\bf c} | \cos \beta.
\end{equation}
After some algebra, 
\begin{equation}
r = \arccos (\cos ^2 \phi).
\end{equation}
Thus we have the angle of incidence, $\theta_i$, required for the calculation
above, and we have the angle, $\beta$, so that the $R_{\parallel}$ and
$R_{\perp}$ factors can be applied to the rotated $E_x$ and $E_y$.

\section{Results and discussion}
\label{sec:problems}

\subsection{Simulation Results}

\begin{figure*}
\begin{center}

\includegraphics[width=6in]{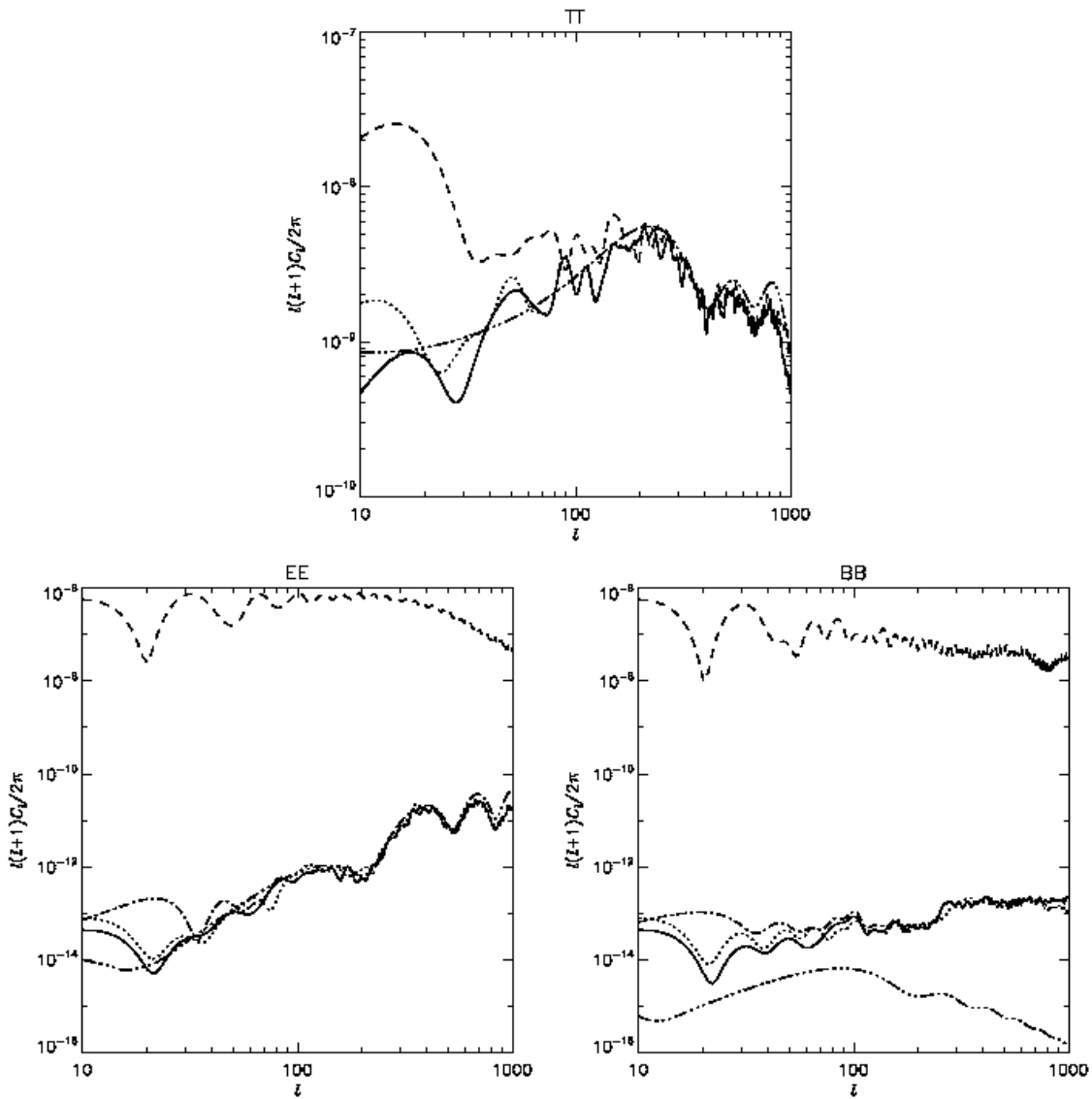}
\caption{TT, EE, and BB spectra from various maps. dash-dot-dot-dot
line is the input spectra, as simulated by CAMB. Dotted line is the
all sky map, masked with the same mask used for the observed
patches. solid is after our pipeline, but with no mirror. dashed is
the spectra after having been observed with a flat fold
mirror. dot-dash is the spectra after observed, but with the fitting
algorthim (described in the text) applied.}
\label{fig:goodflatten}
\end{center}
\end{figure*}

We use CAMB to simulate a sky, observe it with our scanning strategy
code, and follow the pipeline through to calculate power
spectra. Firstly we test our pipeline by ignoring the effect of the
fold mirror; resultant APS are shown in
figure~\ref{fig:goodflatten}. In the region of interest
($\ell\sim$100--1000) the output spectra match the input spectra well
for TT and EE. The spectra from the masked pipeline maps for TT and EE
are slightly supressed for larger $\ell$; 5\% at at $\ell\sim 900$
when compared to the spectra made from the full sky maps. However, as
the spectra from the masked full sky maps are also supressed, we
conclude that this is due to the effect of the partial sky mask and
fairly naive apodization procedure. Similarly, the BB spectra from the
masked map is higher than the all-sky map, but as the pipeline BB
spectra agrees with the masked all-sky map BB spectra, we conclude
that it is the mask rather than the map-making procedure which is
increasing the BB power. Although this level of analysis is not
sufficient for a full B-mode experiment, this level of analysis is
sufficient for investigating the effect of the flat-fold mirror.

\begin{figure}
\includegraphics[angle=-90,width=3in]{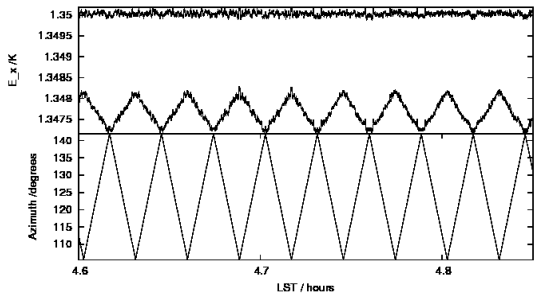}
\caption{Upper plot: time stream of the input and observed $E_x$ for a
  noiseless simulation, with a $28.2\times 10^{-9} \Omega {\rm m}$
  fold mirror. Lower plot: time stream of the telescope azimuth.}
\label{fig:timestream}
\end{figure}

We then apply the effect of a flat fold mirror in the pipeline, using
the resistivity of aluminium of $28.2\times10^{-9}$~$\Omega {\rm m}$,
and observations at 90GHz.  A section of the time stream is shown in
figure~\ref{fig:timestream}, both the input and output $E_x^2$ and the
azimuth of the observed sky location. The correlation between induced
polarization with observed sky azimuth is clear.

\begin{figure}
\includegraphics[width=3in]{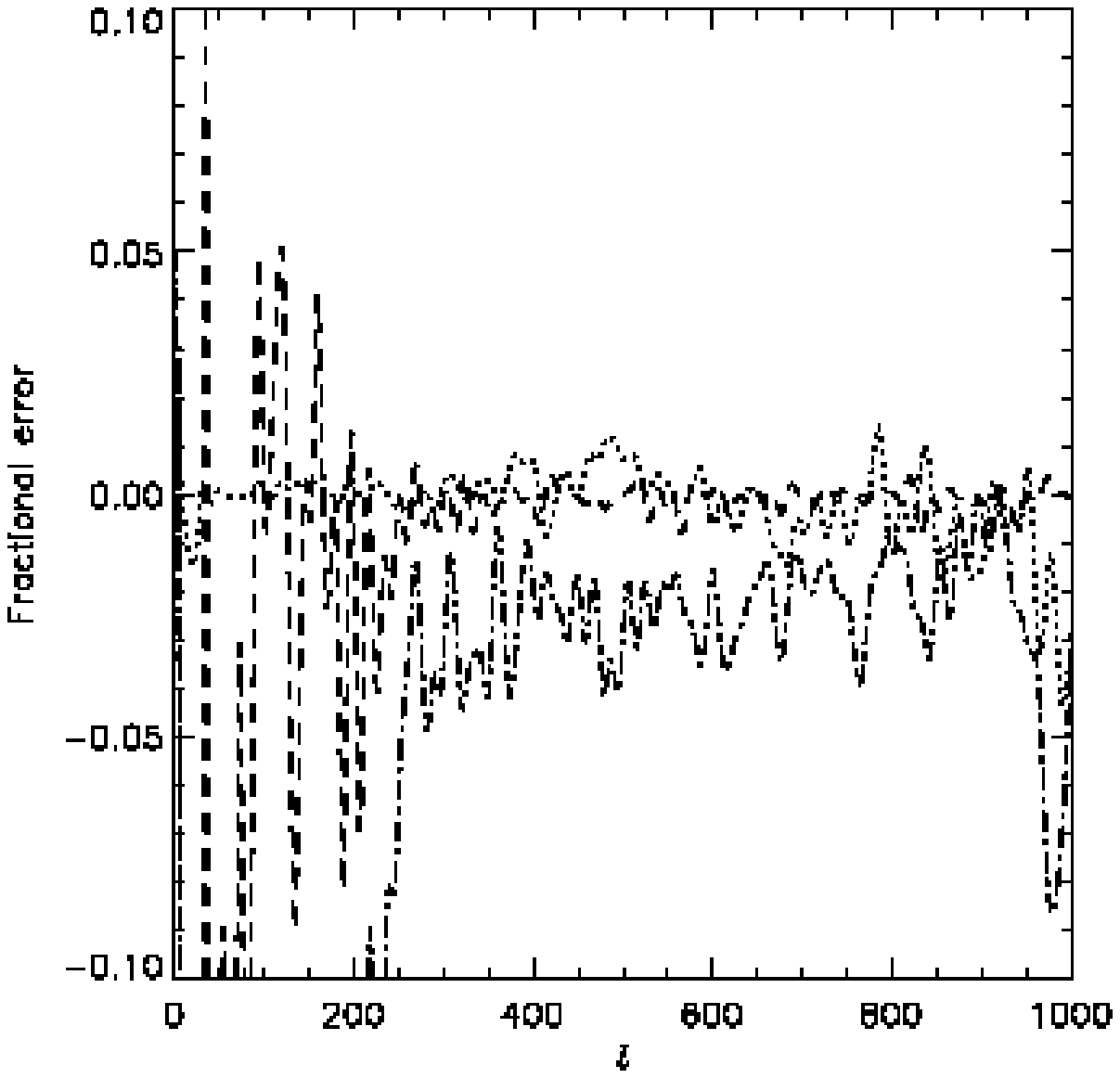}
\caption{Fractional errors for the TT (dotted), EE (dashed) and BB
(dot-dash) spectra. Calculated as the difference between the no-mirror
case and the fitted mirror case, divided by the no-mirror case.}
\label{fig:spectraratios}
\end{figure}

The induced polarization is big, and so it is reasonable to remove it
without requiring the entire data set. By applying the reverse of the
observing algorthim, we can determine a ``pre-mirror'' $E_x$ and $E_y$
for a given mirror resitivity. We use $\Sigma (E_x^2 - E_y^2)$, summed
over the day of observation for a particular pixel, as a measure of
the amount of polarization, and minimize it by varying the resistivity,
using the Nelder-Mead simplex direct search \cite{matlabminimizer},
implemented in Matlab. This time-line
can then be re-inserted into the pipeline, maps made and spectra
computed. The resultant APS are shown in
figure~\ref{fig:goodflatten}. We see that in the range $\ell\sim$
100--1000, the TT, EE and BB spectra all match the pipeline output as
before. In order to show problems more clearly, we also plot the
difference between the spectra produced from the maps with and without
the flat fold mirror. These are shown against $\ell$ in
figure~\ref{fig:spectraratios}.  We find that the TT and EE spectra
are reconstructed well, with less than 1\% difference between
$\ell$ of 250 and 900. The BB spectra shows both fluctuation of $\sim
1\%$ and an overall offset of 2.5\% over the same $\ell$ range.

%

\subsection{Noise}

The shortest required observing time, for a generalized CMB
experiment, would be achieved using a photon noise limited telescope
at a good site. For a instrument such as Clover, situated in the
Atacama desert, the system NET is 175$\mu$K~$\sqrt{\rm s}$. With a
100Hz data stream, this noise level overwhelms the induced
polarization signal from T $\to$ Q,U mixing.  However, simple binning
by a factor of ten is sufficient to see the induced polarization
signal in the presence of white noise. If we add white noise at the
level of 175 $\mu$K~$\sqrt{\rm s}$, with an $\rho = 28.2 \times
10^{-9}$~$\Omega {\rm m}$, the best fitting resistivity returned with
this procedure is $28.8\times 10^{-9}$~$\Omega {\rm m}$.

Detector $1/f$ noise is assumed to be unimportant, assuming a
detector-based modulation strategy is implemented (e.g.\ using a
rotating or stepped half-wave plate such as EBEX, or a fast detector
switching such as QUIET). Atmospheric $1/f$ total power variations
contributes strongly however as the effect of the reflection is a
total power to polarization conversion. Thus the atmospheric $1/f$
becomes a polarization $1/f$ term which does not get removed by any
modulation strategy. A straightforward differencing strategy, for
example differencing successive samples during azimuthal scanning, or
differencing two adjacent detectors, to remove this varying
atmospheric contribution is likely to be the most efficient method,
but would reduce the effective observing time by a factor of two. 

\subsection{Temperature variations}

The resistivity of aluminium varies with temperature, and so the
induced polarization will vary with temperature. Given a resistivity
at room temperature of $\rho_0 = 28.2 \times 10^{-9} \Omega {\rm m}$,
the difference between the two polarization reflection coefficients is
0.075\%. We introduce a temperature coefficient of resistance,
$\rho_T$, of $3.8\times10^{-3} {\rm K}^{-1}$, with $\rho = \rho_0 (1 +
\rho_T \Delta T)$. Varying the temperature by 20$^{\circ}$C gives a
difference of 0.078\% between the two polarization reflection
coefficients, meaning that the T $\to$ Q,U mixing varies by
$\sim7$\%. For a reasonably large flat fold mirror -- in this case the
flat is $\sim 2$m diameter -- the temperature variation is likely to
be on the timescale of hours, driven largely by the ambient
temperature variations. Such variations should be removable by fitting
the resistivity to the data over different timescales on a
pixel-by-pixel basis. We note that, apart from the pixel on the
optical axis, each pixel will see a section of the mirror whose
location varies in a predictable way with azimuth (and elevation) and
so fitting with respect to mirror position might also be required. In
this case, the error on the fit can be reduced by fitting multiple
pixel data simulatiously to produce a mirror resistivity map.

\subsection{Oxide growth}

%


The polarizing effect of the fold mirror comes from its
resistivity. As such, effects that alter the resistivity alter the
polarizing effect. Aluminium grows an oxide layer very quickly, which
then hydrates. The hydrated aluminium oxide can then be degraded
chemically, or by physical damage to the surface. Such a surface is
thin (5-15 nm) and so does not change the resistivity by a large
amount. It is not unreasonable to assume that there will be changes to
the surface during an extended observing campaign, both with oxide
thickness and additional surface contamination. As with the
temperature variation, contamination would almost certainly be
localised, and so fitting resistivity with respect to mirror position
would probably be required. However such changes would be expected to
be slow (timescales of days or more), or at least sudden and stable,
and so more data is available to fit out the effect. As a result, we
expect the resistivity variation from surface oxidation and
contamination to be better characterised than the temperature
variation.


\subsection{Polarized foregrounds}
The work so far has assumed that the light incident on the flat-fold
mirror is polarized at a very low level, and that the majority of the
polarized signal comes from the flat-fold mirror.  This is the case
when considering the CMB only, where the polarization signal has zero
mean.  These assumptions do not hold, however, when observing an
arbitrary patch of sky, where both synchrotron and dust from the
galaxy is expected to be polarized at the $\sim 1-10$\% level, with a
total intensity of 10--100 $\mu$K close to the Galactic Plane, but
down to $\lesssim 1 \mu$K at high Galactic latitudes. As such, we
would recommend this method only be used in regions of known low
foreground emission, where the resulting error in polarization
reconstruction would be small.  Over sufficiently large areas of sky
the magnitude of the effect could be modelled, and in regions with
relatively little large-scale variation in the foreground emission,
the effect would be a constant offset, similar to an additional
absolute calibration error.  The areas of sky observed by most B-mode
experiments are selected to meet these requirements.

In order to test this, we re-ran the simulation with some fiducial dust
and synchrotron models added. After applying the same method as discussed
above, the map-plane residuals are large; we see residuals of +20 to
-40 $\mu$K in the Q map, in comparison with the signal level of $\pm
20 \mu$K across the map. We leave fully developing this method to work
in the presence of significant polarized foregrounds for future work. 

\section{Conclusions}

We have analyzed the effect of a flat-fold mirror on polarization
measurements of the CMB. In the case of no foregrounds and no noise
with no corrections, we find that the E and B spectra are heavily
contaminated by the mixing of total power to polarization
signal. However, as the CMB is expected to have a net polarization of
zero, the effect can be fitted out. In the case of having a net
polarization, such as in a real observation of the sky, the
reconstructed maps contain significant residuals.



\section{Acknowledgements}

We thank Enzo Pascale and Giorgio Savini for useful discussions during
this work. Some of the results in this paper have been derived using
the HEALPix\cite{healpix}. 

\end{document}